\documentclass[aps,prl,onecolumn,nofootinbib,superscriptaddress,notitlepage]{revtex4-1}
\usepackage[a4paper,top=3.00cm,bottom=2.50cm,left=3.00cm,right=3.00cm]{geometry}
\usepackage{amsmath}
\usepackage{amssymb}
\usepackage{enumerate}
\usepackage{mathptmx}
\usepackage{epsf}
\usepackage{rotate}
\usepackage{graphicx}
\DeclareGraphicsExtensions{.eps, .pdf, .jpg, .png}
\usepackage[usenames,dvipsnames]{color}
\usepackage{hyperref}
\usepackage{bookmath}
\usepackage{mathfrak}
\usepackage{mymatrices}
\usepackage{mygroups}
\def\gimt{\hat{g}_{\mbox{\small imt}}}
\begin{document}
%~~~~~~~~~~~~~~~~~~~~~~~~~~~~~~~~~~~~~~~~~~~~~~~~~~~~~~~~~~~~~~~
\title{
\vspace*{-2.0cm}
\hspace*{-20mm}{\small {\sf Chapter in the Proceedings of the 1996 Verditz Workshop on}}
{\small {\sl Quasiclassical Methods in Superconductvity and Superfluidity}}
\vspace*{-0.05cm}
\\
\hspace*{-115mm}{\small {\sf Edited by D. Rainer and J.A. Sauls (1998)}}
\vspace{1.5cm}
\\
Impurity States in D-wave Superconductors
}
\author{D. C. Chen}
\affiliation{Department of Physics \& Astronomy, Northwestern University, Evanston, IL 60208} 
\author{D. Rainer}
\affiliation{Physikalisches Institut, Universit\"at Bayreuth, D-94550 Bayreuth, Germany}
\author{J.A. Sauls}
\affiliation{Department of Physics \& Astronomy, Northwestern University, Evanston, IL 60208}

\begin{abstract}
The structure of the order parameter and the excitation spectrum are investigated for isolated impurities in d-wave superconductors. Atomic scale impurities, or defects, scatter quasiparticles and lead to local suppression (pair-breaking) near the impurity. The pair-breaking effect arises from the formation of quasiparticle states bound to the impurity. The corresponding reduction in spectral weight in the pair condensate is responsible for pair-breaking. The formation of the bound state is due to multiple Andreev scattering by the combined effects of potential scattering, which leads to changes in momentum of the scattered quasiparticle, and the anisotropy of the d-wave order on the Fermi surface. The spectral weight of the bound state decays exponentially away from the impurity on a length scale $\xi_{*}=\hbar v_f/\sqrt{|\Delta(\vp_f)|^2-\eps_{*}^2}$, where $\eps_{*}$ is the energy of the impurity state. The continuum spectrum exhibits Tomasch oscillations due to the interference between Andreev reflected particle- and hole-like quasiparticles.
\end{abstract}

\maketitle

%~~~~~~~~~~~~~~~~~~~~~~~~~~~~~~~~~~~~~~~~~~~~~~~~~~~~~~~~~~~~~~~

\vspace*{-10mm}
\section{Introduction}

There is growing evidence that superconductivity
in many of the high T$_c$ cuprates is described by
an {\it unconventional}
condensate of spin-singlet pairs with an orbital pairing amplitude
having ``d-wave'' symmetry, or more precisely $d_{x^2-y^2}$ symmetry.
In unconventional superconductors the order parameter,
$\Delta(\vp_f)$, is strongly anisotropic on the Fermi surface,
and in general breaks the symmetry of the crystal point group.
The discoveries of novel superconducting properties in strongly 
correlated systems such as liquid \He, high T$_c$ and heavy Fermion metals,
and more recently superfluid \He\ confined 
in aerogel (Thuneberg (this volume)),
have motivated extensive theoretical investigations
into the thermodynamic and transport properties of unconventional
superconductors. Among the key questions are the effect of
scattering by impurities, grain boundaries and surfaces on the
superconducting properties. It is well known that
non-magnetic impurity scattering is
a strong pairbreaker for unconventional superconductors.
In this article we examine the local structure of
the order parameter and excitation 
spectrum in the neighborhood of an atomic impurity
within the d-wave model for the high T$_c$ cuprates.
The combined effects of potential scattering and 
the broken reflection symmetry of the d-wave order
parameter leads to the formation of a localized
state bound to the impurity by multiple Andreev 
scattering. The electronic structure near an impurity
located on the surface of a superconductor can be studied with 
atomic resolution by scanning tunneling microscopy (STM). 
Other techniques
like NMR and scanning SQUID spectroscopy offer additional local 
tools for investigating the local electronic and magnetic environment
near atomic defects.

Theoretical investigations into the structure of impurities in
systems with unconventional pairing were developed in detail for
superfluid \He\ beginning with the study of the currents
induced around an impurity in \Hea\ by \cite{rai77}.
The condensation energy loss from an ion in
\Heb\ was calculated by \cite{thu81}, and the
quasiparticle bound-state spectrum for hard-sphere ions in 
\Heb\ calculated by \cite{thu81b}. Moving ions in \Heb\
lead to emission of quasiparticles, which are responsible for
the drag force of an ion in superfluid \He. The nonequilbrium
distribution of quasiparticles emitted by a moving ion in
\Heb\ was calculated in quasiclassical theory by \cite{ash88}.
The quasiclassical theory developed for the study of impurities in 
superfluid \He\ has been used to study
the current and magnetic field distribution around a non-magnetic
impurity in unconventional pairing models of heavy Fermion 
superconductors with broken time-reversal symmetry \cite{cho89a}. 

The possibility of unconventional pairing in the high T$_c$ 
supercondutors has led to further interest in the 
properties and electronic structure of impurities in
unconventional superconductors
\cite{cho94,bal95,bal96}. In particular,
Choi studied the anisotropy of the Tomasch oscillations
in the local density of states (DOS) \cite{tom65}
near an isolated impurity, while
Balatsky and co-workers \cite{bal95,bal96}
focussed on the impurity bound state, or resonance, for an
impurity in a 2D $d_{x^2-y^2}$ superconductor.

\section{Quasiclassical Theory}

We investigate the properties
structure and excitation spectrum of an impurity in a two-dimensional
metallic plane of a $d_{x^2-y^2}$ superconductor.
Our approach follows closely the theory developed in the
early 80's for ions in superfluid \He. 
We calculate changes in the superconducting properties
resulting from scattering by an atomic impurity to quasiclassical 
accuracy. We start from Eilenberger's 
transport equation \cite{eil68},
\be\label{Eilenberger-Matsubara}
\hspace*{-2mm}
\left[i\eps_n\hat{\tau}_3-\hat{\Sig}(\vp_f,\vR;\eps_n)-\hat{\Delta}(\vp_f,\vR)\,,
 \hat{g}(\vp_f,\vR;\eps_n)\right]+
 i\vv_f\cdot\grad_{\vR}\hat{g}(\vp_f,\vR;\eps_n)=0
\,,
\ee
for the matrix propagator, $\hat{g}(\vp_f,\vR;\eps_n)$.
For non-magnetic scattering in spin-singlet superconductors
we can ignore the spin degree of freedom and work within the 
$2\times 2$ particle-hole space of the matrix propagator,
\be
\hat{g}(\vp_f,\vR;\eps_n)=
g\,\hat\tau_3+f\,\hat\tau_+\,+\,\bar{f}\,\hat\tau_-
%\left(\matrix{g & f \cr \bar{f} & \bar{g}}\right)
\,,
\ee
where $\hat\tau_3$, $\hat\tau_{\pm}=(\hat\tau_1\pm i\hat\tau_2)/2$
are the Pauli matrices in particle-hole space.
The diagonal propagator determines observable quantities,
\eg the equilibrium current density,
\be
\vj(\vR)=2N_f\int\,d\vp_f\,e\vv_f\,
         T\sum_{\eps_n}\,g(\vp_f,\vR;\eps_n)\,,
\ee
as well as the spectral density of excitations
at the position $\vR$ for states near $\vp_f$ on the Fermi surface,
\be
N(\vp_f,\vR;\eps)=-\frac{1}{\pi}\,\Im\,g^{R}(\vp_f,\vR;\eps)
\,.
\ee
The retarded propagator is obtained by analytic continuation;
$g^R(\eps)=g(i\eps_n\rightarrow\eps+i0)$.
The off-diagonal propagator is related to the pair-amplitude and
determines the pairing self-energy (``order paramater'') through
the BCS \cite{bar57} gap equation,
\be
\Delta(\vp_f,\vR)=\int\,d\vp_f'\,V(\vp_f,\vp_f')\,
T\sum_{\eps_n}^{|\eps_n|<\epsilon_c}\,f(\vp_f',\vR;\eps_n)
\,,
\ee
where $V(\vp_f,\vp_f')=V_d\,\eta(\vp_f)\eta^*(\vp_f')$ is the pairing 
interaction in the
weak-coupling limit for pairs of quasiparticles with relative
momenta $\vp_f$ scattering into pair states with relative
momenta $\vp_f'$; $\epsilon_c$ is the BCS cutoff which is
eliminated with the pairing interaction in favor of the 
measured transition temperature, $\ln(1.13\epsilon_c/T_c)=1/V_d$.
We also assume only one pairing channel is relevant; for $d_{x^2-y^2}$
pairing the basis functions, $\eta(\vp_f)$, transform like
$(\hat{p}_x^2-\hat{p}_y^2)$;
the order parameter then has the form,
$\Delta(\vp_f,\vR)=\Delta(\vR)\,\eta(\vp_f)$. We can eliminate the
pairing interaction and cutoff from the gap equation and write,
\be
\ln(T/T_c)\Delta(\vR)=\int\,d\vp_f\,\eta(\vp_f)\,
T\sum_{\eps_n}\,\left(f(\vp_f',\vR;\eps_n)
-\frac{\pi\Delta(\vp_f,\vR)}{|\eps_n|}\right)
\,.
\ee

Eilenberger's equation is a differential equation
for the quasiclassical propagator, $\hat{g}(\vp_f,\vR;\eps_n)$, 
along {\it classical trajectories} defined by the Fermi
velocity, $\vv_f$, for excitations in the normal state.
The self-energy terms ($\hat\Sig$ and $\hat\Delta$) 
in the transport equation represent
forces acting of the excitations described
by the quasiclassical matrix distribution function. These
forces arise from external sources, \eg coupling
to the electromagnetic field, and from the coupling of quasiparticles
to the medium, \eg Landau's molecular field and the electron-phonon
interaction. The pairing self-energy that enters the
transport equation is off-diagonal in particle-hole space.
This term produces coherent mixing of particle and hole
excitations which is responsible for most of the novel 
coherence effects observed in superconductors.

Below we consider the effects of scattering by an
impurity atom on the excitation spectrum and order parameter of
a layered d-wave superconductor.
An atomic scale impurity in a metal is a strong,
short-range interaction. It cannot be treated
as a perturbative correction to electronic wavefunctions;
however, a theory for impurity scattering
in a superconductor can be formulated within the
quasiclassical framework \cite{thu81}
with a small expansion parameter,
$\sigma/\xi_0$, where
$\sigma$ is the (linear in 2D) cross-section of an impurity 
for scattering of normal-state quasiparticles at the
Fermi surface, and $\xi_0=v_f/2\pi T_c$ is the
superconducting coherence length. This ratio is typically
small in low T$_c$ superconductors and superfluid \He, however
it may be of order $0.1$--$0.5$ in high T$_c$ superconductors.

The interaction of quasiparticles
in a superconductor with an impurity enters the
quasiclassical theory as {\it source term} on the
right-hand side of Eilenberger's
transport equation \cite{thu81},
\be\label{Impurity-Source}
\hat{I}(\vp_f;\eps_n)=
\delta(\vR-\vR_{\mbox{\small imp}})\,
\left[\hat{t}(\vp_f,\vp_f;\eps_n)\,,\,
\hat{g}_{\mbox{\small imt}}(\vp_f;\eps_n)\right]
\,,
\ee
where the $\hat{t}$-matrix desribes multiple
scattering by an impurity located at $\vR_{\mbox{\small imp}}$,
\ber
\hat{t}(\vp_f,\vp_f',\vR_{\mbox{\small imp}};\eps_n) =
\hat{u}(\vp_f,\vp_f')
+ N_f\int d\vp_f''\,\hat{u}(\vp_f,\vp_f'')\,
\hat{g}_{\mbox{\small imt}}(\vp_f'',\vR_{\mbox{\small imp}};\eps_n)\,
\hat{t}(\vp_f'',\vp_f',\vR_{\mbox{\small imp}};\eps_n)
\,,
\eer
and $\hat{g}_{\mbox{\small imt}}$ is the {\it intermediate propagator} 
describing propogation of excitations in the absence of the impurity;
$\gimt$ is obtained by solving the Eilenberger's equation without
the impurity source term, but in general with the self-consistently determined
order parameter in the presence of the impurity. Like the full
propagator the intermediate propagator
satisfies the normalization condition, $\gimt^2=-\pi^2\hat{1}$.

The interaction of quasiparticles with an impurity enters as the
matrix element of an
effective potential, $u(\vp_f,\vp_f')$, which scatters a quasiparticle
at $\vp_f\rightarrow\vp_f'$ on the Fermi surface. We
consider the simplest model for the impurity; non-magnetic potential
scattering described by a single ``s-wave'' scattering amplitude, $u_0$.
The normal-state $t$-matrix can be expressed either in terms
of the interaction potential, or the scattering phase shift,
\be
t^R_N=\frac{u_0}{1+i\pi N_f u_0}=
    \left(\frac{-1}{\pi N_f}\right)\sin\delta_0\,e^{i\delta_0}
\,,
\ee
where $\delta_0=\tan^{-1}\left(-\pi N_f u_0\right)$ is the s-wave phase shift.
The corresponding scattering cross-section (in $2D$) of the impurity is
$\sigma=(2\pi\hbar/p_f)\,\bar{\sigma}$, where
\be
\bar{\sigma}=\sin^2\delta_0=\frac{(\pi N_f u_0)^2}{1+(\pi N_f u_0)^2}
\,,
\ee
is normalized to the cross-section in the unitarity limit
($u_0\rightarrow\infty$), \ie twice the Fermi wavelength.

The source term is localized at the impurity site and gives a
significant correction to the propagator at a point near the
impurity for the trajectories that intersect the impurity.
We can calculate the leading correction to the propagator and order 
parameter as an expansion in $\sigma/\xi_0$. In this limit
the corrections are expected to be small, \eg
$\delta\Delta\sim (\sigma/\xi_0)\Delta_0$, so we can approximate
$\gimt$ by the bulk propagator,
\be
\gimt\simeq\hat{g}_{\mbox{\small bulk}}=
-\pi\frac{i\eps_n\hat{\tau}_3-\hat{\Delta}(\vp_f)}
         {\sqrt{\eps_n^2+|\Delta(\vp_f)|^2}}
\,,
\ee
and expand the full propagator about the bulk solution,
\be
\hat{g}(\vp_f,\vR;\eps_n)=g_{\mbox{\small bulk}}(\vp_f,\vR;\eps_n)
+\delta\hat{g}(\vp_f,\vR;\eps_n)
\,.
\ee
We can then Fourier transform the linearized transport equation for
the leading order correction, $\delta\hat{g}$, to obtain
\be
\left[
i\eps_n\hat{\tau}_3-\hat{\Delta}(\vp_f)\,,
\delta\hat{g}(\vp_f,\vq;\eps_n)
\right]
+\vv_f\cdot\vq\,\delta\hat{g}(\vp_f,\vq;\eps_n)=
\hat{S}(\vp_f,\vq;\eps_n)
\,.
\ee
The new source term,
\be\label{Impurity-Source+Vertex}
\hspace*{-2mm}\hat{S}(\vp_f,\vq;\eps_n)=
\left[\delta\hat{\Delta}(\vp_f,\vq)\,,\,
\hat{g}_{\mbox{\small bulk}}(\vp_f;\eps_n)\right]
+
\left[\hat{t}(\vp_f,\vp_f;\eps_n)\,,\,
\hat{g}_{\mbox{\small bulk}}(\vp_f;\eps_n)\right]
\,,
\ee
includes the first-order correction in $\sigma/\xi_0$
to the order parameter, $\delta\hat{\Delta}$.

The linearized transport equation can be solved efficiently with
the aid of the normalization conditions \cite{rai94b}, from
which we obtain,
\ber\label{linear-response}
\delta\hat{g}&=&\,\,\onehalf\,
\left(i\eps_n\hat{\tau}_3\,-\hat\Delta(\vp_f)\,+\onehalf\vq\cdot\vv_f\,\hat{1}\right)^{-1}\,
\hat{S}(\vp_f,\vq;\eps_n)
\nonumber
\\
&=&-\onehalf\,
\frac{\left(i\eps_n\hat{\tau}_3\,-\hat\Delta(\vp_f)\,-\onehalf\vq\cdot\vv_f\,\hat{1}\right)}
{\eps_n^2 +| \Delta(\vp_f) |^2 + \onefourth\left(\vv_f\cdot\vq\right)^2}\,\,\,\,\,\,
\hat{S}(\vp_f,\vq;\eps_n)
\,.
\eer
We can calculate the asymptotic corrections to
the order parameter and the local DOS in the 
region $R\gg\sigma$ of the impurity, and compare these results
with self-consistent numerical solutions
of the transport equations and the gap equation discussed below.
For the numerical solutions we use the ``explosion method'' 
described by Kurkij\"arvi 
(this volume) and the optimized relaxation method of M\"obius 
and Eschrig \cite{eschrig89} to solve the transport equations
and self-consistent self-energy equations.

\section{Local Structure of the Order Parameter}

\begin{figure}
\includegraphics[width=0.465\textwidth]{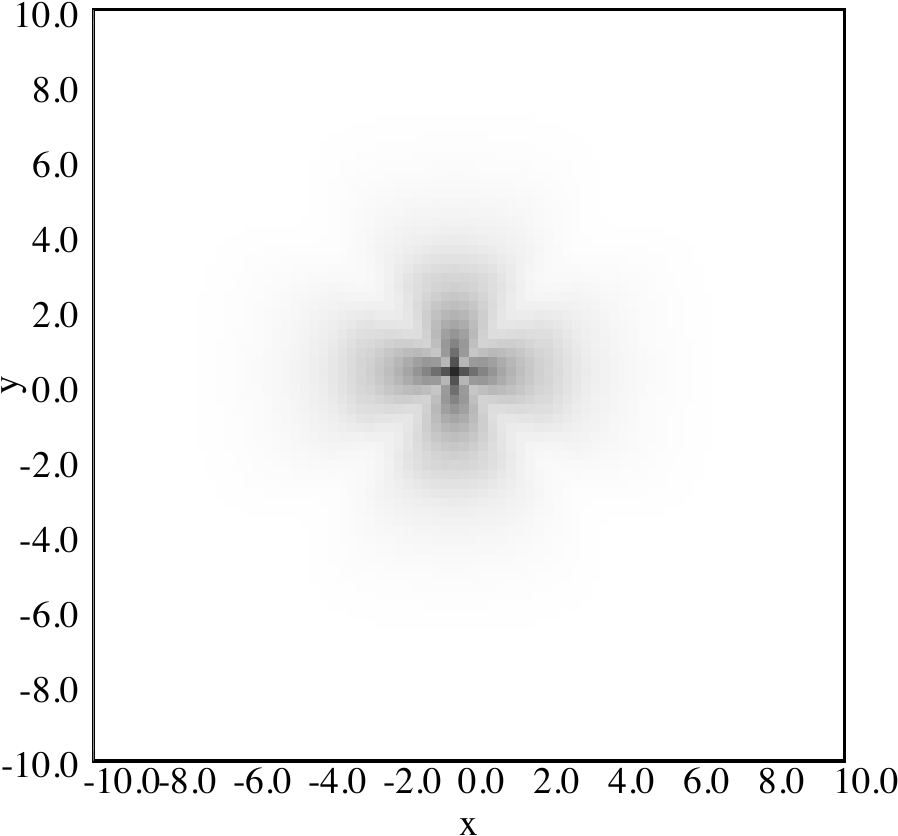}
\includegraphics[width=0.45\textwidth]{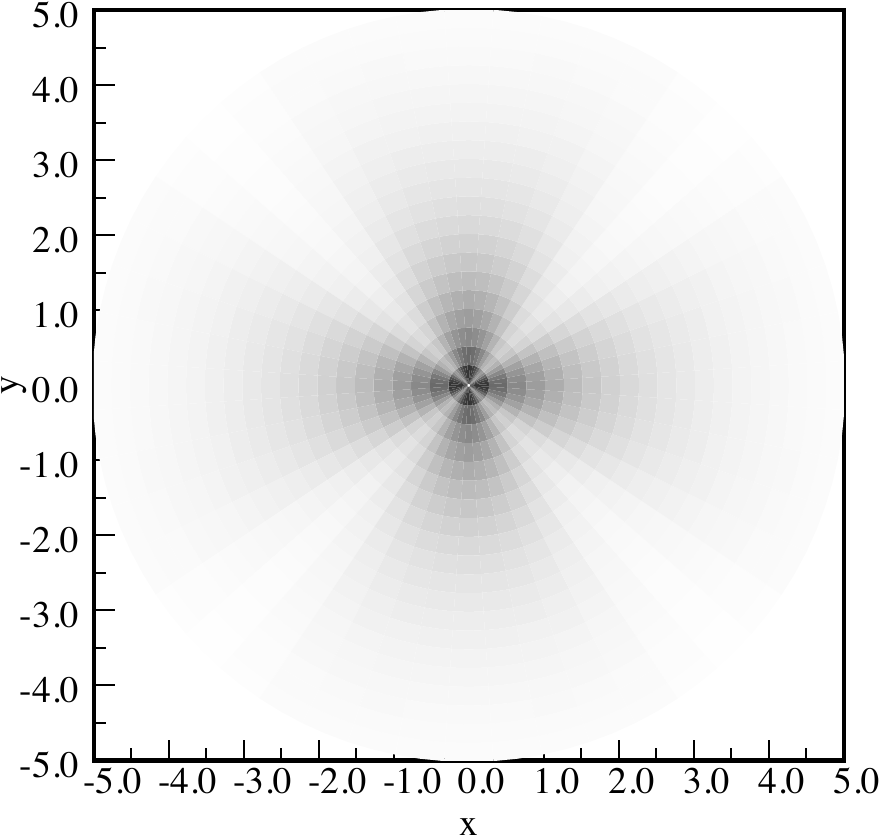}
\caption{\small Structure of the order parameter near an isotropic 
impurity in a $d_{x^2-y^2}$ superconductor. 
(a) cross-section $\bar{\sigma}=0.1$, 
(b) cross-section $\bar{\sigma}=0.5$. The space coordinates
are scaled in units of $\xi_0$.
}
\label{fig-OP_structure}
\end{figure}

In the asymptotic region, $\sigma\ll R \ll \xi_0$, the correction
to the order parameter is small, of order $\sigma/R$,
is dominated by the trajectory that intersects the impurity and
decays as $1/R$ away from the impurity. The
asymptotic pairbreaking suppression to the 
order parameter, obtained from
Eq.(\ref{linear-response}) and the gap equation,
scales as,
\be
\delta\Delta(\vR)\approx -\frac{\sigma}{R}\,|\eta({\hat\vR})|^2\,,
\ee
in the unitarity limit.
Self-consistent numerical calculations shown in Fig.~\ref{fig-OP_structure}
exhibit the d$_{x^2-y^2}$ anisotropy of the order parameter.
Note that it is the singular contribution from the trajectory 
with $\hat{\vR}=\hat{\vp}_f$ in 
the limit $q\sim 1/R\gg 1/\xi_0$ that
determines the anisotropy of the order 
parameter in position space.

%~~~~~~~~~~~~~~~~~~~~~~~~~~~~~~~~~~~~~~~~~~~~~~~
%~~~~~~~~~~~~~~~~~~~~~~~~~~~~~~~~~~~~~~~~~~~~~~~

\section{Bound States and Particle-Hole Coherence}

In general the broken reflection symmetry of the $d_{x^2-y^2}$
order parameter enforces nodes of the excitation gap on the Fermi
surface. The rate at which the gap opens at the nodal
points is not fixed by symmetry, but by the microscopic pairing 
interaction and the Fermi surface. In the limit where the nodal
regions of the Fermi surface are small enough that they can be 
neglected we can approximate the order parameter by 
``nodeless'' basis functions,
\be
\Delta(\vp_f,\vR)=\Delta(\vR)\,
    \mbox{sgn}(\hat{\vp}_x^2-\hat{\vp}_y^2)
\,,
\ee
with $d_{x^2-y^2}$ symmetry; \ie the order parameter
changes sign across the (110) reflections planes.
This model is used below because
it allows one to separate the effects associated with
the low-energy nodal quasiparticles
from the effects due to the broken reflection
symmetry of the order parameter, \eg the
suppression of the order parameter by impurity scattering.
The pairbreaking effect on the order parameter is closely related to
the appearance of quasiparticle states below the gap edge,
$|\Delta(\vp_f)|$. Most of the spectral weight lost from the condensate
appears in the single particle spectrum as a bound state, or resonance.
The properties of the quasiparticle bound state and the continuum
near an impurity are discussed below for the nodeless d-wave model.

\begin{figure}
\includegraphics[width=0.475\textwidth]{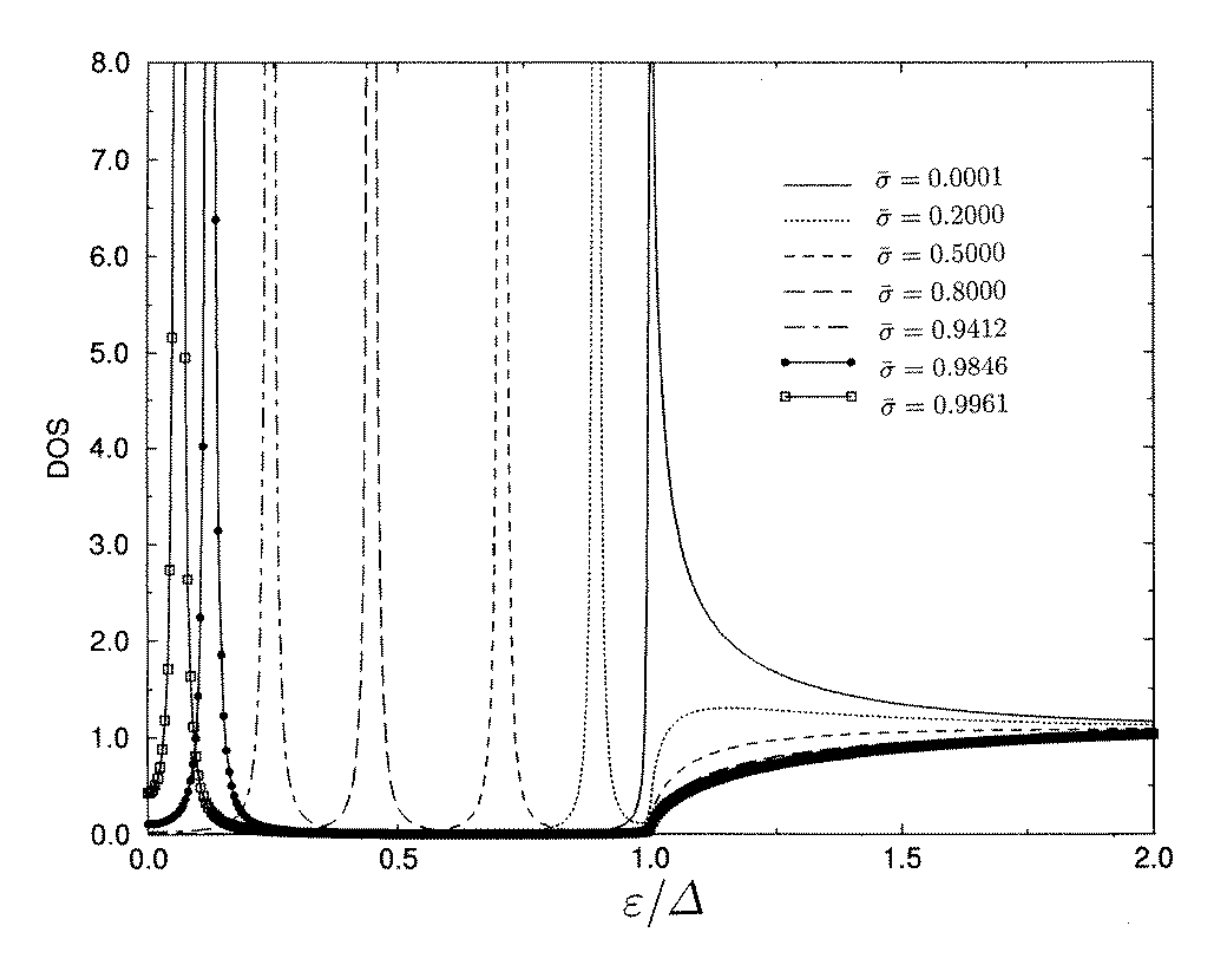}
\includegraphics[width=0.465\textwidth]{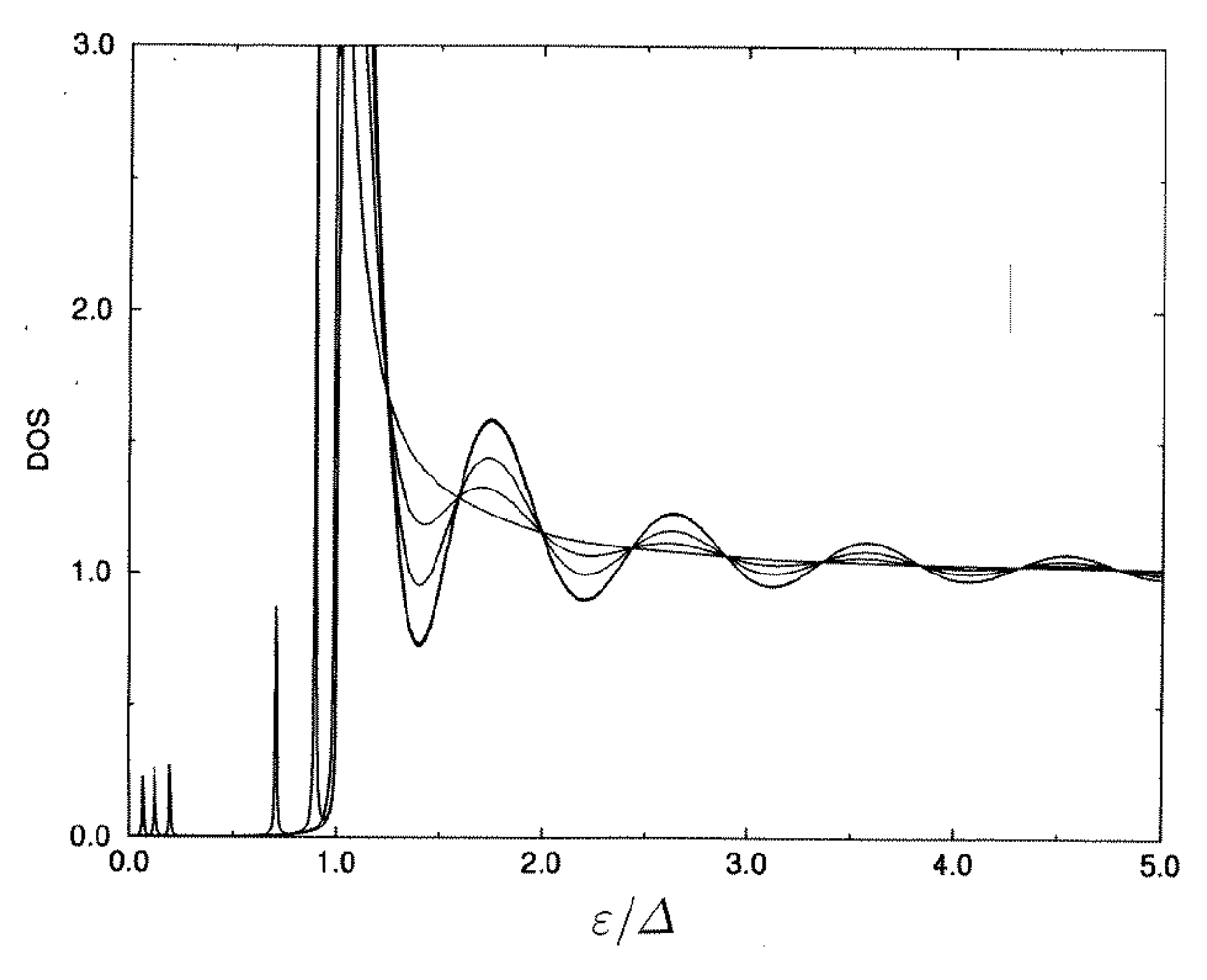}
\caption{\small Local DOS near an impurity as a function of energy 
for various cross sections.
(left) DOS at $R=0$ showing the bound state 
for various values of $\bar{\sigma}$.
(right) DOS at $R=10\,\xi_0$ for 
$\bar{\sigma}=0.01$, $0.2$, $0.5$, $0.96$, $0.98$, $0.996$.
The Tomasch oscillations increase in amplitude for increasing
cross section.
}
\label{fig-DOS_bound-states}
\end{figure}

In Fig.~\ref{fig-DOS_bound-states} (left panel) we show results for the local density of states 
for the nodeless d-wave model
as a function of the impurity cross section, $\bar{\sigma}$,
ranging from the Born limit ($\bar{\sigma}\ll 1$)
to the strong scattering limit ($\bar{\sigma}\approx 1$). The left panel
is the local DOS at the position of the impurity, while the
right panel is the local DOS far from the impurity, $R=10\,\xi_0$.
In the unitarity limit the bound state occurs at the Fermi level;
the spectral weight near the gap edge is removed and appears at the 
Fermi level as a sharp zero-energy bound state. For weaker scattering
the bound state moves to higher energy and has reduced spectral
weight which is shifted to the continuum; in the Born limit
the bound state  merges into the contiuum and we recover the 
the familiar BCS singularity in the DOS. 

%~~~~~~~~~~~~~~~~~~~~~~~~~~~~~~~~~~~~~~~~~~~~~~~
%~~~~~~~~~~~~~~~~~~~~~~~~~~~~~~~~~~~~~~~~~~~~~~~

The impurity-induced quasiparticle bound state
appears as a pole in the $\hat{t}$-matrix. For s-wave
impurity scattering and a $d_{x^2-y^2}$ pairing state we can
evaluate the $\hat{t}$-matrix to leading order in $\sigma/\xi_0$
by replacing $\gimt$ with the bulk propagator and order
parameter. Thus, $\hat{t}$ reduces to
\be
\hat{t}(\eps_n)=u_0\hat{1}
   +N_f\,u_0\,\langle\hat{g}_{\mbox{\small bulk}}\rangle\,\hat{t}(\eps_n)
\,,
\ee
where $\langle\hat{g}_{\mbox{\small bulk}}\rangle$ is the Fermi surface
average of the bulk propagator, which reduces to
$\langle\hat{g}_{\mbox{\small bulk}}\rangle =
\hat{\tau}_3\,\int d\vp_f\,$
$i\eps_n/\sqrt{\eps_n^2+|\Delta(\vp_f)|^2}$
because of the vanishing of the Fermi surface average,
$\langle\hat{\Delta}(\vp_f)\rangle=0$, for a d-wave order parameter,
a feature which is common to many unconventional pairing models.
The $\hat{t}$-matrix reduces to the form
$\hat{t}= t_1\,\hat{1} + t_3\,\hat{\tau}_3$. Only the
$\hat{\tau}_3$ term contributes to the impurity source term
and is given by
\be
t_3(\eps_n)=
\frac{1}{\tau_N}
\frac{i\eps_n\sqrt{\eps_n^2+|\Delta|^2}}{\eps_n^2+|\Delta|^2(1-\bar{\sigma})}
\,,
\ee
for the nodeless d-wave model, with $1/\tau_N=(1/\pi N_f)\bar{\sigma}$ 
being the normal-state transport scattering rate.
The $\hat{t}$ matrix has a pole on the real axis below the gap at an energy,
$\eps_{*}=\sqrt{1-\bar{\sigma}}\Delta$.
Since the $\hat{t}$-matrix is the source for the linear response propagator 
the bound state pole appears in the single particle excitation
spectrum.

The right panel of
Fig.~\ref{fig-DOS_bound-states} shows more clearly the shift in spectral weight between
the bound state and the continuum as a function of the scattering
cross-section. In this case the spectral weight in the
bound state is lower for stronger scattering. This is because the
radial position of the bound state wave function also moves towards 
the impurity as the binding energy increases. 
Thus, for $R=10\,\xi_0$ the peak in the wavefunction moves towards
the point $R$ as the cross-section decreases, and the spectral weight 
of the bound state increases with decreasing cross-section.
Away from the impurity the local DOS shows oscillations of
fixed period (in energy) which increase in amplitude with
increasing cross-section (right panel of Fig.~\ref{fig-DOS_bound-states}).
These are Tomasch oscillations
resulting from the interference of particle- and hole-like
quasiparticles undergoing impurity-induced Andreev scattering. 

%~~~~~~~~~~~~~~~~~~~~~~~~~~~~~~~~~~~~~~~~~~~~~~~
\begin{figure}
\includegraphics[width=0.90\textwidth]{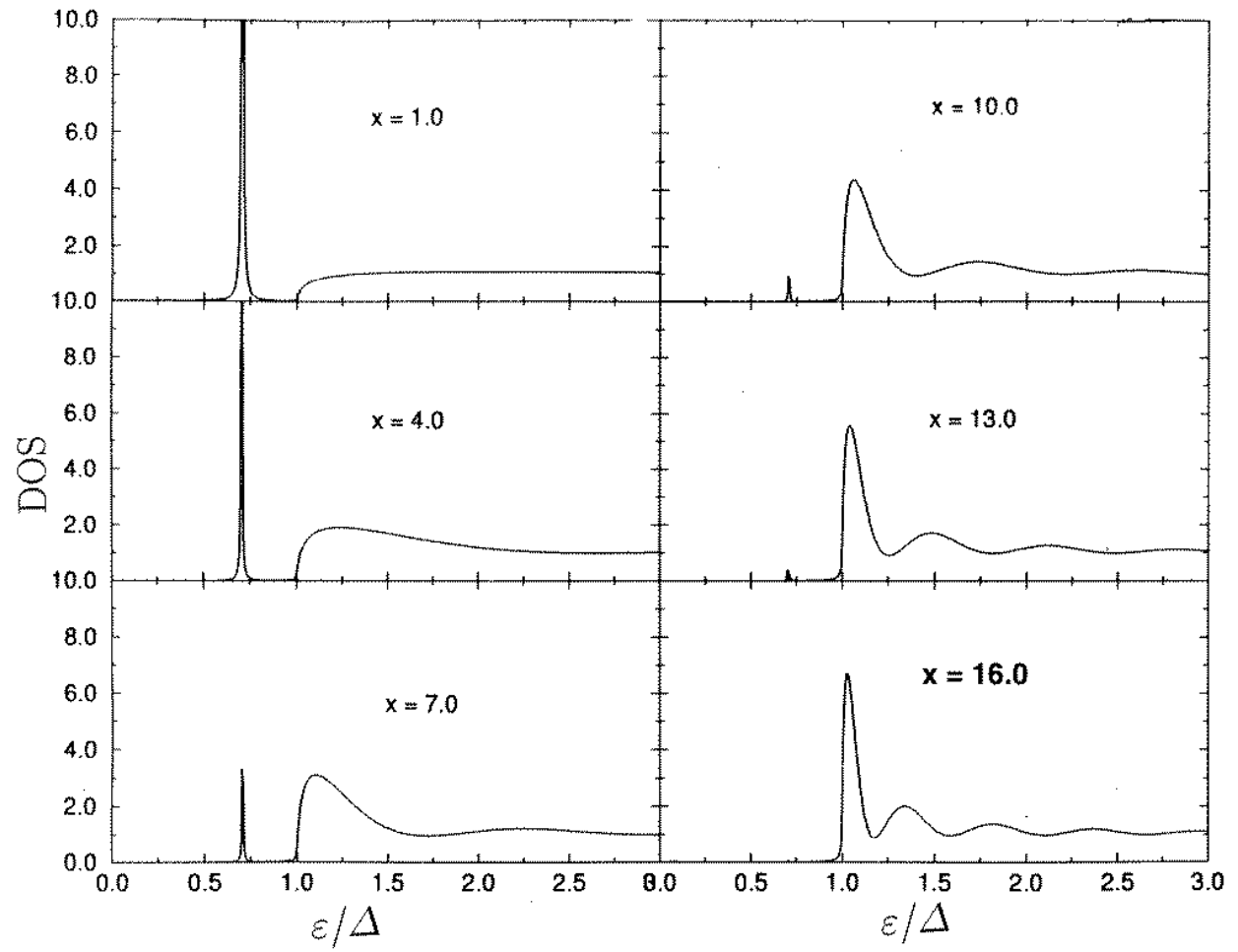}
\caption{\small Local DOS near an impurity as a function of energy for 
various distances from the impurity
for an intermediate scattering cross section, $\bar{\sigma}=0.5$. The bound
state appears at $\eps_{*}\simeq 0.7\Delta$. The spectral weight
shifts to the continuum with increasing distance from the impurity
and Tomasch oscillations develop for $R\gg\xi_0$ with a wavelength
given by $\lambda_T(\eps)\simeq 3.5\xi_0/\sqrt{(\eps/\Delta)^2-1}$.
}
\label{fig-DOS_Tomasch_oscillations}
\end{figure}
%~~~~~~~~~~~~~~~~~~~~~~~~~~~~~~~~~~~~~~~~~~~~~~~

Figure~\ref{fig-DOS_Tomasch_oscillations} shows the spectrum as a 
function of distance from the
impurity for a fixed scattering cross section of $\bar{\sigma}=0.5$ and
bound state energy of $\eps_{*}\simeq 0.7\Delta$.
The spectral weight in the bound state peak measures the square of the
bound state wavefunction as a function of position, 
\be
|\Psi(\vR)|^2\propto
\int_{\eps_{*}-\Delta\eps}^{\eps_{*}+\Delta\eps}\,d\eps\,N(\vR;\eps)
\sim\frac{1}{R}e^{-R/\xi_{*}}
\,,
\ee
where the range of the bound state scales as $\xi_{*}=\hbar
v_f/\sqrt{|\Delta|^2-\eps_{*}^2}\simeq\xi_0/\sqrt{\bar{\sigma}}$.
The integration over the local DOS is restricted to the 
region of the bound state peak.
At distances $R\agt 4\xi_0$ Tomasch oscillations develop in the
continuum with wavelength 
$\lambda_{T}(\eps)=\hbar v_f/\sqrt{\eps^2-|\Delta|^2}$

Both the continuum oscillations and the
bound state result from ``particle-hole coherence'',
which is a fundamental signature of BCS superconductivity,
and is responsible for essentially all
non-classical phenomena associated with superconductivity.
Andreev reflection is a particulary striking example, 
and is the origin of the Tomasch oscillations and the
impurity bound state. Andreev reflection is a quantum-mechanical
process of branch conversion in which a quasiparticle 
(with group velocity parallel to the Fermi momentum) 
is reflected into a hole with the same momentum
but opposite group velocity. 
Scattering from non-magnetic surfaces or impurities 
has no effect on the equilibrium order parameter or
excitation spectrum in conventional isotropic 
superconductors \cite{and59,abr59a}.
The absence of pairbreaking by non-magnetic scattering
can be traced to the fact that such scattering is
not associated with Andreev scattering. 
However, in unconventional superconductors
in which the order parameter changes sign on the Fermi surface,
potential scattering of quasiparticles around the Fermi surface
induces Andreev scattering. Bound states and peaks in the continuum
spectrum result from constructive interference of multiply
scattered particle- and hole-like quasiparticles.

This work was supported by the NSF through the Science and Technology
Center for Superconductivity (DMR 91-20000).

%--------------------------------------------------------------------
%\bibliographystyle{apsrev4-1_PRX_style.bst}
%\bibliography{CM,QFS,Books}
%--------------------------------------------------------------------
%
\end{document}